\documentclass[12pt]{article}

\usepackage{graphicx}
\def\pt{$p_T$ }
\def\dis{distribution}

\def\bq{\begin{equation}}
\def\eq{\end{equation}}

\begin{document}

\vskip1cm

\begin{center}  {\Large {\bf Particles Associated with $\Omega$ 
Produced \\ at Intermediate $p_T$}}
\vskip .75cm
   {\bf Charles B.\ Chiu$^1$ and  Rudolph C. Hwa$^{2}$}
\vskip.5cm

   {$^1$Center for Particle Physics and Department of Physics\\
University of Texas at Austin, Austin, TX 78712, USA\\
\bigskip
$^2$Institute of Theoretical Science and Department of
Physics\\ University of Oregon, Eugene, OR 97403-5203, USA}
\end{center}

\begin{abstract}
The dual observation of the $\Omega$ production in central Au-Au 
collisions having both an exponential \pt \dis\ and also associated 
particles above the background has been referred to as the $\Omega$ 
puzzle. We give a quantitative description of how that puzzle can be 
understood in terms of phantom jets, where only ridges without peaks 
are produced to give rise to both the $\Omega$ trigger and its 
associated particles. In the framework of recombination of thermal 
partons we are able to reproduce both the $\Delta\phi$ \dis\ and the 
trigger-momentum dependence of the yield of the associated particles. 
We make predictions on other observables that can be checked by 
further analyses of the data.
\end{abstract}
\newpage
\section{Introduction}

Recent data on the production of $\phi$ and $\Omega$ at the 
Relativistic Heavy ion Collider (RHIC) at intermediate  $p_T$  have 
revealed important properties of the $s$ quarks in the dense medium 
created by the collision of Au nuclei at 200 GeV \cite{ba,ja}.  Both 
$\phi$ and $\Omega$ exhibit exponential behavior in their $p_T$ 
distributions up to $p_T  \approx 4.5$ GeV/c.  Such properties of the 
$\phi$ and $\Omega$ spectra have been reproduced in \cite{hy} as 
consequences of recombination of thermal partons.  The implication is 
that before hadronization the dense system consists of thermalized 
partons that include the $s$ quarks.  That is the main characteristic 
that quark-gluon plasma (QGP) should possess.  The production of 
$\phi$ and $\Omega$ provides a clear window through which one can 
observe the thermal source without the contamination of shower 
partons due to hard scattering, which is suppressed for $s$ quarks. 
Since QGP is a description of the bulk medium, one can reasonably ask 
what the properties are at the edge of that medium, namely, when the 
partonic $k_T$ is $> 1$ GeV/c.  Recent data from STAR on correlation 
that uses $\Omega$ as trigger in the intermediate $p_T$  region has 
revealed interesting properties of the associated particles 
\cite{jbs}.  The dual features of the exponential spectrum of 
$\Omega$ and the existence of associated particles above background 
have been referred to as the $\Omega$  puzzle \cite{rch}.  This paper 
is aimed at providing a quantitative resolution of that puzzle.

Particles produced at intermediate and high $p_T$  in heavy-ion 
collisions are associated with jets due to hard scattering of 
partons.  The medium effect on hadronization has been successfully 
described in the recombination model in terms of the recombination of 
thermal and shower partons \cite{hy2}.  The single-particle $p_T$ 
distributions of $\pi$, $K$ and $p$ are well reproduced, showing 
power-law deviation from the exponential behavior at low \pt that is 
due to the recombination of thermal partons.  The hadronization 
mechanism that works for those particles that have light quarks as 
constituents does not work for the production of $\Omega$ that 
contains only $s$ quarks.  Strange shower partons are suppressed 
whatever the initiating hard parton may be \cite{hy}.  For that 
reason the $p_T$  distribution of $\Omega$ is essentially exponential 
up to the highest measured value of nearly 6 GeV/c \cite{ja2}, as has 
been well reproduced by the recombination of three thermal $s$ quarks 
\cite{hy}.

Since shower $s$ quark does not participate in the formation of 
$\Omega$, it is natural to conclude that jets play no role and that 
the production of $\Omega$ is not accompanied by any associated 
particles, which are usually present in events triggered by less 
strange hadrons.  The distributions of associated particles in the 
latter case have also been satisfactorily described by the 
recombination of thermal and shower partons in $p_T$  \cite{hy3}, as 
well as in $\Delta \eta$ and $\Delta \phi$ \cite{ch}.  Without jet 
production there should be no jet structure characterized by 
associated particles distinguishable from the background.

The discovery that $\Omega$ trigger is accompanied by associated 
particles above the background has been astounding \cite{jbs,lr,ob}. 
How can a trigger particle formed by thermal partons have partners 
above the background, since they are presumably also of thermal 
origin?  That is the $\Omega$ puzzle.

\section{The Ridge}

Conventional jet structure consists of a peak above a pedestal, when 
the trigger momentum ($p_T ^{\rm  trig}$) and those of the associated 
particles ($p_T ^{\rm  assoc}$) are not too high.  STAR has adopted 
the terminology Jet (J) for peak and ridge (R) for pedestal.  The J 
to R ratio of their yields depends on $p_T ^{\rm  trig}$, $p_T ^{\rm 
assoc}$ and centrality \cite{jp}.  For $p_T ^{\rm  trig} > 4$ GeV/c 
and $p_T ^{\rm  assoc} > 2.3$ GeV/c in central Au-Au collisions at 
200 GeV, it has been shown that $J/R {\stackrel{<}{\sim}} 1$. 
However, at lower values of $p_T ^{\rm  assoc}$ the $J/R$ ratio is 
smaller, becoming as low as $0.1 - 0.15$ at $p_T ^{\rm  assoc}\sim 
1.2$ GeV/c \cite{jb2}.   For $\Lambda$-triggered events that ratio is 
even lower $(<0.1)$.  No data for that ratio yet exist for 
$\Omega$-triggered events, but one may anticipate it to be further 
lower according to the trend of increasing strangeness.  If that is 
true, then the particles associated with $\Omega$ will not exhibit 
any significant peak in the $\Delta \eta$ distribution, which is 
where J and R are usually seen with non-strange triggers.  On the 
basis of the study done in \cite{hy}, where $\Omega$ is shown to 
originate from thermal sources, we can assert that the Jet component 
should be absent, and that only the ridge will be observed.

The notion of phantom jet was suggested in \cite{rch} to describe the 
phenomenon in which a jet is produced without the Jet part being 
observed.  The ridge that is observed provides the evidence that 
there is an underlying jet.  Such a scenario has not been confirmed 
by data, but is our conjecture as a solution to the $\Omega$ puzzle, 
the quantitative description of which will be given in the following 
two sections.  A confirmation would be in evidence when the 
$\Omega$-triggered $\Delta \eta$ distribution exhibits a ridge 
without a peak.  At this point the statistics in the data for $2.5 < 
p_T ^{\rm  trig} < 4.5$ GeV/c and $1.5 < p_T ^{\rm  assoc} < p_T 
^{\rm  trig}$ is not high enough to give a discriminating $\Delta 
\eta$ distribution.  What is seen in the $\Delta \phi$ distribution 
cannot distinguish Jet from ridge because there is no elongation in 
the $\phi$ direction as there is in the $\eta$ direction due to 
longitudinal expansion \cite{ch}.

A phantom jet is seen as an ordinary jet if the trigger is a 
non-strange particle.  It may be initiated by a gluon or a light 
quark, which can generate non-strange shower partons that hadronize 
as $\pi$, $K$, $p$ or $\Lambda$.  If $p_T ^{\rm  trig}$ is not too 
high, the hard scattering may occur in the interior of the bulk 
medium.  The scattered hard parton traverses the medium and loses 
energy on its way out to the surface.  The energy deposited in the 
medium enhances the thermal motion of the partons in the vicinity of 
the trajectory.  Those partons hadronize and form the ridge that is 
above the background.  Such a sequence of processes is conventional, 
and has been described successfully in \cite{ch} that gives both the 
$\Delta \eta$ and $\Delta \phi$ distributions.  However, the 
$\Omega$-triggered events are notconventional, even though the 
underlying jet is conventional as described above.  The reason is 
that the $s$ quarks are suppressed in the shower.  Without the $s$ 
shower partons $\Omega$ cannot be formed as a direct consequence of 
hard scattering.  It can, however, be formed indirectly from the 
ridge that is thermal and has ample supply of $s$ quarks.  Events 
triggered by $\Omega$ produced that way are rare at high $p_T ^{\rm 
trig}$, since the thermal spectrum is exponential.  Thus the $p_T 
^{\rm  trig}$ threshold is set low, like 2.5 GeV/c.  The 
corresponding $p_T ^{\rm  assoc}$ threshold is even lower, like 1.5 
GeV/c, in order to have enough statistics to produce a meaningful 
associated particle distribution.  When $p_T ^{\rm  assoc}$ is so 
low, the ridge component dominates over the peak so the J/R ratio is 
very small.  The ordinary jet thus becomes a phantom jet under the 
condition of low $p_T ^{\rm  trig}$ and lower $p_T ^{\rm  assoc}$.

If the ridge can supply $s$ quarks to form the $\Omega$ trigger, it 
can surely supply light quarks to produce other lower-mass charged 
hadrons.  Those are the associated particles detected \cite{jbs}. 
Thus the ridge of the phantom jet is the key to the solution of the 
$\Omega$ puzzle.  The $\Omega$ has a $p_T$ distribution that is 
exponential, on the one hand, because it is formed by the thermal 
partons in the ridge, and it has associated particles, on the other 
hand, because the ridge is above the background.

\section{The Background}

Our aim now is to describe the $\Delta \phi$ distribution of the 
associated particles in events triggered by $\Omega$ \cite{jbs}.  The 
signal is less than 4\% of the background height.  Thus in order for 
us to reproduce quantitatively the signal, it is necessary to show 
first that the background can be accurately obtained in our 
formalism.  There is a $v _2$ oscillation due to elliptic flow, 
resulting in \cite{jbs}
\begin{eqnarray}
{dN^{\rm bg}\over d\Delta\phi} = 6.9 + 0.1 \cos 2 \Delta \phi
\label{1}
\end{eqnarray}
for the background.  We aim to get the central value of 6.9 from the 
single-particle distribution of pion for $p_T < 3$ GeV/c that we have 
obtained previously \cite{hy2,ch}.

The experimental conditions are that in 0-10\% central Au-Au 
collisions at 200 GeV/c the $\Omega$ trigger is detected in the 
window $2.5 < p_T ^{\rm  trig} < 4.5$ GeV/c with $|\eta|< 1$ and the 
unidentified charged particles associated with it are in the range of 
$1.5 <  p_T ^{\rm  assoc}  < p_T ^{\rm  trig}$.  Let the trigger 
distribution be denoted by
\begin{eqnarray}
N(1 )= {dN_\Omega\over dp_1 d\eta_1}
\label{2}
\end{eqnarray}
and the two-particle distribution by
\begin{eqnarray}
N(1,2)={dN_{\Omega,2}\over dp_1dp_2 d\eta_1 d\Delta\eta d\Delta\phi} \ ,
\label{3}
\end{eqnarray}
where $\Delta \eta = \eta_2 - \eta _1$,  $\Delta \phi = \phi_2 - \phi 
_1$, and $p_i$ being short for $p_{iT}$.  The associated particle 
distribution per trigger is then
\begin{eqnarray}
{dN \over d\Delta\eta d\Delta\phi} = {\int_{p_c}^{p_d} dp_1 
\int_{p_a}^{p_1} dp_2 \int_{-1}^{1} d\eta_1 N(1,2) \over 
\int_{p_c}^{p_d} dp_1   \int_{-1}^{1} d\eta_1 N(1)} \ ,
\label{4}
\end{eqnarray}
where $p_a = 1.5$, $p_c = 2.5$ and $p_d = 4.5$ all in units of GeV/c. 
For the background distribution $N(1, 2)$ is factorizable, i.e.,
\begin{eqnarray}
N^{\rm bg}(1,2) = N(1)N^{\rm bg}(2)\ .
\label{5}
\end{eqnarray}
It follows that on the right hand side of Eq.\ (4) the $\eta_1$ integrations in the numerator and the denominator are trivially cancelled, leaving
\begin{eqnarray}
{dN^{\rm bg}\over d\Delta\eta d\Delta\phi} = {\int_{p_c}^{p_d} dp_1 
N(p_1) N^{\rm bg} (p_1, \Delta\eta, \Delta\phi) \over 
\int_{p_c}^{p_d} dp_1 N(p_1)},
\label{6}
\end{eqnarray}
where
\begin{eqnarray}
N^{\rm bg} (p_1, \Delta\eta, \Delta\phi) = \int_{p_a}^{p_1} dp_2 
{dN^{\rm bg} \over dp_2 d\Delta\eta d\Delta\phi}.
\label{7}
\end{eqnarray}
Since $|\eta|< 1$, $\Delta \eta$ ranges from $-2$ to $+2$. Assuming $dN^{bg}/dp_2d\Delta\eta\Delta\phi$ 
is independent of $\Delta\eta$,  we obtain upon integration of 
(\ref{7}) over $\Delta \eta$
\begin{eqnarray}
N^{\rm bg} (p_1, \Delta\phi) =4 \int_{p_a}^{p_1} dp_2 {dN^{\rm bg} 
\over dp_2 d\Delta\eta  d\Delta\phi}\equiv  \int_{p_a}^{p_1} dp_2 
{dN^{\rm bg} \over dp_2  d\Delta\phi}.
\label{8}
\end{eqnarray}
For 0\% centrality there is no azimuthal asymmetry, so the $\Delta 
\phi$ distribution of the background is uniform.

We can determine $dN^{\rm bg}/dp_2$ from our previous work on 
recombination model  \cite{hy2,ch} so as to establish a connection 
with that formalism, on the basis of which we shall calculate the 
ridge in the next section.  The particles in the background are 
formed by the recombination of thermal partons, whose invariant 
distribution is \cite{hy2}
\begin{eqnarray}
{\cal T}_0 (q) = q {dN^{\rm th} \over dq} =  C_0 q e^{-q/T_0}  ,
\label{9}
\end{eqnarray}
where
\begin{eqnarray}
C_0 = 23.2 \, ({\rm GeV/c})^{-1}, \qquad T_0 = 0.317 \,  {\rm GeV/c}.
\label{10}
\end{eqnarray}
The recombination of those partons to form thermal pions is given by \cite{hy2}
\begin{eqnarray}
{dN^{\rm th}_{\pi} \over dp} = {1 \over p^2} \int^{p}_{0} dq \, {\cal 
T}_0 (q) {\cal T}_0 (p-q) = {C^2_0 \over 6} pe^{-p/T_0},
\label{11}
\end{eqnarray}
where the recombination function $x_1 x_2 \delta (x_1 + x_2 - 1)$, 
$x_i$ being the momentum fraction, has been used in the integral 
above.  That is for any one of the three charge states of pions.  For 
the charged pions we gain a factor of 2.  For other charged 
particles, we take $K^{\pm}/\pi ^{\pm} = b$ and $p^{\pm}/\pi ^{\pm} = 
cp$, where $p$, as for all momentum symbols, denotes transverse 
momentum.  The numerical factors can be taken to be $b = 0.4$ and $c 
= 0.3$ (GeV/c)$^{-1}$ as reasonable approximations of the data, where 
the linear growth in $p$ of the ratio $p^{\pm}/\pi ^{\pm}$ is 
approximately valid up to about 3 GeV/c
  \cite{hy2}.  Thus for unidentified charged particles in the 
background we adopt for the integrand in Eq.\ (\ref{8}) the form
\begin{eqnarray}
{dN^{\rm bg} \over dp_2 d \Delta \phi} = {4 \over 3} C^2_0 p_2 (1 + b 
+ cp_2)e^{-p_2/T_0},
\label{12}
\end{eqnarray}
Upon integration over $p_2$ from $p_a$ to $p_1$, we obtain
\bq
N^{bg} (p_1,  \Delta \phi) = {4 \over 3} \, (C_0 T_0)^2 h(p_1, T_0), \label{13}
\eq
where
\bq
h(p_1, T_0)\equiv
f\left(\frac{p_a}{T_0},T_0\right)- f\left(\frac{p_1}{T_0},T_0\right), 
\label{14}
\eq
with
\bq
  f(x,T_0)= (1 + b) (1 + x) + cT_0 (2 + 2x + x^2)e^{-x}.
\label{15}
\eq

To carry out the integrations over $p_1$ in Eq.(\ref{6}) we need the 
single-particle distribution for the production of $\Omega$, which 
can be determined in the same way as for proton by recombination 
\cite{hy,hy2}, except that it is simpler when only thermal $s$ quarks 
contribute.  The result is \cite{hy}
\begin{eqnarray}
{dN_{\Omega} \over dp_1} = {C^3_{\Omega}\over 27}{p^3_1 \over 
p_{10}}e^{-p_1/T_s},
\label{16}
\end{eqnarray}
where $p_{10} = \left(p^2_1 + M^2_{\Omega} \right)^{1/2}$ and
\begin{eqnarray}
T_s = 0.33\ {\rm GeV/c}.
\label{17}
\end{eqnarray}
The normalization factor $C_{\Omega}$ is immaterial, since it appears 
in both the numerator and the denominator of Eq.(\ref{6}), so they 
cancel each other.

  Integrating both over $p_1$ and  $\Delta\eta$ in Eq.\ (\ref{6}), we 
finally have
\begin{eqnarray}
{dN^{\rm bg}\over  d\Delta\phi} = {4 \over 3} (C_0T_0)^2 \left< 
h(p_1, T_0)\right>,
\label{18}
\end{eqnarray}
where
\begin{eqnarray}
  \left< h(p_1, T_0)\right> =
\int^{p_d}_{p_c} dp_1 \,
\frac{dN_{\Omega}} {dp_1}\, h (p_1, T_0) \left/
  \int^{p_d}_{p_c} dp_1 \frac{dN_{\Omega}} { dp_1}\right.
\equiv
f\left(\frac{p_a}{T_0},T_0\right)- 
\left<f\left(\frac{p_1}{T_0},T_0\right)\right>.
\label{19}
\end{eqnarray}
The last term on the right hand side of Eq.(19) is given by
\begin{eqnarray}
\left<f(\frac{p_1}{T_0},T_0)\right>
=\frac{1}{ I(x_s,y_s)}\frac{T_{0s}^3}{T_s^3 }
\int_{x0s}^{y0s} dx
\left[(1+b)(1+z)+\tau_0(2+2z+z^2)\right]
\frac{x^3}{(x^2+\mu_{0s}^2)^{1/2}}\,e^{-x}, \label{20}
\end{eqnarray}
with the variable of integraton $x=p_1/T_s$. The limits of 
integration are $x_{0s}=p_c/T_{0s}$ and $y_{0s}=p_d/T_{0s}$, where 
$T_{0s}=T_0T_s/(T_0+T_s)$. The other symbols are $z= (T_{0s}/T_0)x$, 
$\tau_0=cT_0$ and $\mu_{0s}=m_\Omega/T_{0s}$. The denominator of the 
first factor is given by
\begin{equation}
I(x_s,y_s)=\int_{x_s}^{y_s} dx \frac{x^3}{(x^2+\mu_s^2)^{1/2}}e^{-x}, 
\label{21}
\end{equation}
with the integration variable $x=p_1/T_s$. Other symbols $x_s$, $y_s$ 
and $\mu_s$ are respectively $p_c/T_s$, $p_d/T_s$, and 
$m_{\Omega}/T_s$.

For the trigger momentum range being specified by $p_c$ and $p_d$ 
given just below Eq.\ (\ref{4}), the effective value of $p_1$ found 
for $\left< h(p_1, T_0)\right>$ is 2.8 GeV/c.  Using the parameters 
for the thermal partons given in Eq.\ (\ref{10}), we can now compute 
the background height for the associated particles and obtain
\begin{eqnarray}
{dN^{\rm bg}\over  d\Delta\phi}  = 6.9.
\label{22}
\end{eqnarray}
This result agrees well with the data \cite{jbs}, and gives us 
confidence in the formalism to proceed to the calculation of the 
ridge.

It should be noted that the $\Omega$ spectrum is stated in Eq.\ 
(\ref{16}) without reference to it being in the ridge, since the 
calculation was done in \cite{hy} using a fitted value of $T_s$.  The 
fact that the value of $T_s$ being slightly higher than $T_0$ is a 
hint that the thermal source is enhanced.  But the procedure of 
calculating the hadron distribution by recombination of thermal 
partons is insensitive to whether the thermal source is enhanced or 
not.

\section{Associated Particles}

Since our resolution of the $\Omega$ puzzle is based on the validity 
of the assertion that both the trigger and its associated particles 
are formed from the ridge, we must now show that the $\Delta\phi$ 
distribution of the associated particles from that source agrees with 
the data \cite{jbs}.  The calculation of the ridge follows 
essentially the procedure used in \cite{ch}, except that we are 
hampered this time by the lack of data on $\Delta \eta$ distribution 
that was available for unidentified charged triggers.  Indeed, to be 
sure that one has a ridge, it is usually necessary to find it in the 
$\Delta\eta$ distribution that exhibits the longitudinal elongation. 
Since we are dealing with phantom jets that show imperceptible 
evidence for Jet in the events triggered by $\Omega$, all that have 
been seen in $\Delta\phi$ (and presumably will be seen in 
$\Delta\eta$) are in the ridge.

The basic difference between the ridge and the background is that for 
the former we have new parameters $C$ and $T$ for the enhanced 
thermal partons, while for the latter $C_0$ and $T_0$ are given in 
Eq.\ (\ref{10}).  There is also a $\Delta\phi$ dependence which is 
forced to vanish at $\Delta\phi = \pm 1$ due to the subtraction 
procedure used in the data analysis.  We write
\begin{eqnarray}
C(\Delta\phi) = C_0 H(\Delta\phi) = C_0H_0 \exp \left[ - 
(\Delta\phi)^2/2\sigma^2\right],
\label{23}
\end{eqnarray}
where the values of $H_0$ and $\sigma$ are discussed below.  Since 
associated particles in the ridge are unidentified charged hadrons, 
just as in the background, we obtain by thermal recombination as done 
in the preceding section
\begin{eqnarray}
{dN^R \over  d\Delta\phi} = {4 \over 3} (C_0T)^2 H (\Delta\phi) 
\left< h(p_1, T)\right> - {dN^{\rm bg}\over  d\Delta\phi},
\label{24}
\end{eqnarray}
where the last term is as calculated in Eq.\ (18), given in (22), but 
supplemented by $v_2$ oscillation and therefore behaves as in Eq.\ 
(\ref{1}).

The value of $T$ for the enhanced thermal source may be inferred from 
either $T_s$ or the value of $\Delta T = T - T_0 = 15$ MeV determined 
in \cite{ch}.  They are consistent with each other.  We shall use $T 
= T_0+0.015= 0.332$ GeV.  The width $\sigma$ is fixed by the 
condition $dN^R/d\Delta\phi = 0$ at $\Delta\phi= \pm 1$; it turns out 
to be around 3.6.  We do not know the normalization $C$ of the 
enhanced thermal partons. Even if we did, the width of the ridge in 
$\Delta\eta$ is unknown, since no data exist. The normalization of 
the first term of Eq.\ (\ref{24}) should depend on the width of that 
ridge. At this point the data on the $\Delta\phi$ distribution, being 
blind to the ridge in $\Delta\eta$, are determined by integration 
over the range $-2<\Delta\eta<2$. Thus there is no alternative but to 
fit the height of the peak in $\Delta\phi$ by adjusting the value of 
$H_0$, on which $dN^R/d\Delta\phi$ depends sensitively because Eq.\ 
(\ref{24}) gives the difference between two terms of comparable 
magnitudes.  The data on $dN^R/d\Delta\phi$ are not the only data we 
must fit:  Ref.\ \cite{jbs} also shows the near-side yield for 
various trigger hadrons that all exhibit similar dependence on 
$p^{\rm trig}_T$.  In contrast to $dN^R/d\Delta\phi$ which involves 
integration over $p_1$, the yield $Y(p_1)$ involves integration over 
$\Delta\phi$ instead.  If we denote
\begin{eqnarray}
\tilde{H} = \int^1_{-1}d \Delta\phi H (\Delta\phi) ,
\label{25}
\end{eqnarray}
then the yield is
\begin{eqnarray}
Y(p_1) = {4 \over 3}C^2_0 \left[\tilde{H}T^2 h (p_1, T) - 2T^2_0 h 
(p_1, T_0)\right] .
\label{26}
\end{eqnarray}
Equations (24) and (26) are, of course, interrelated, but they 
provide independent checks on $H_0$ and $T$, since $\left< h(p_1, 
T))\right>$ depends on the range of $p^{\rm trig}_T$, while 
$\tilde{H}$ depends on the range of $\Delta\phi$.

We show in Figs.\ 1 and 2 the results of our calculation for two 
values of $H_0$:
\begin{eqnarray}
(a)\,\, H_0 = 0.795, \qquad (b)\,\, H_0 = 0.790.
\label{27}
\end{eqnarray}
The corresponding values of $\sigma$ are 3.5 and 3.7, respectively. 
For less than 1\% variation in $H_0$ there is a 20\% variation in the 
height of the ridge.  In either case the agreement with data in the 
two figures is good, given the fact that the data have large errors. 
Since the process in which the energy loss by a hard parton is 
converted to the enhanced thermal energy of the medium is not 
calculable, especially since the energy and location of the 
originating phantom jet is unknown, the result that we have obtained 
is perhaps the most one can expect as a phenomenological explanation 
of why the $\Omega$  trigger is accompanied by associated particles 
observable above the background.

Since our calculation of the $\Delta\phi$ distribution and the yield 
does not depend crucially on the exact nature of the trigger, it may 
be taken to explain why our calculated results are roughly in accord 
with the data not only for $\Omega$ trigger, but also for $\Lambda$ 
and $\Xi$ triggers, shown also in Figs. 1 and 2.  Although the $J/R$ 
ratio for $\Lambda$ is not zero, being less than 0.1 would still 
imply that the ridge is dominant, so its associated particles would 
arise mainly from the ridge, with shower partons playing a minimal 
role.  The case with $\Xi$ trigger would be even closer to that of 
the $\Omega$ trigger.

Assuming that the range of the ridge in $\Delta\eta$ is from
  $-2$ to +2, as we have done in Eq.\ (24) in accordance to the 
present experimental cut, we can calculate the total number of 
charged particles in the ridge corresponding to an interval $I$ in 
the $\Delta\phi$ distribution
  \bq
N^R_I=\int_{-I/2}^{I/2} d\Delta\phi \frac{dN^R}{d\Delta\phi}. \label{28}
\eq
  Thus the predicted ridge height (to be compared to the data after 
the removal of the background which includes the $v_2$ contribution) 
is given by
\begin{equation}
\frac{N^R_I}{\Delta\eta}=\frac{N^R_I}{4}=\frac{1}{3} 
C_0^2\left[T^2\hat{H}_I\left<h(p_1,T)\right>-T_0^2I\left<h(p_1,T_0)\right>\right], 
\label{29}
\end{equation}
where $\hat{H}_I=\int_{-I/2}^{I/2} d\Delta\phi H(\Delta\phi)$.
The predicted ridge heights for the two cases considered, for $I=1$, are
\begin{equation}
  (a)\,\, \frac{dN^R}{d\Delta\eta}=0.065, \qquad  (b)\,\, 
\frac{dN^R}{d\Delta\eta}=0.054.   \label{30}
\end{equation}
If the data on the ridge in $\Delta\eta$ turn out to have a width 
different from 4, then the height will also be different accordingly. 
Subject to that qualification, a height of 0.06 is our first-order prediction. 
It is of interest to note that the ridge height envisioned here is 
roughly the same as the one observed in \cite{ja3} for unidentified 
charged trigger with $4<p_T^{\rm trig}<6$ GeV/c and associated 
particles in $2<p_T^{\rm assoc}<4$ GeV/c.

\section{Conclusion}

We have presented a quantitative description of how the $\Omega$ 
puzzle can be resolved. We have shown that the particles associated 
with $\Omega$ can be understood as products of recombination of 
thermal partons in the ridge. Although the non-perturbative dynamical 
process that leads to the formation of the ridge cannot be 
calculated, there are aspects of the data that need coordinated 
explanation in a specific hadronization scheme. We have reproduced in 
Figs.\ 1 and 2 both the $\Delta\phi$ \dis\ and the $p_T^{\rm trig}$ 
dependence of the yield of the associated particles. One parameter 
$H_0$ is used to fit the height of $dN^R/d\Delta\phi$, which depends 
on the exact value of $H_0$ so sensitively (to within 1\% in the 
margin of error) that it is beyond the scope of any dynamical theory 
to predict. What we have learned from this phenomenological study is 
that the particles in the peak observed in the $\Delta\phi$ \dis\ are 
all from the ridge, contrary to the usual identification of peaks 
with Jets. Phantom jets have no peaks. Our finding implies that there 
should not be a peak in the $\Delta\eta$ \dis, still to be produced 
by further analysis of the data. When that \dis\ becomes available 
and shows only a ridge, then our solution of the $\Omega$ puzzle will 
finally be considered confirmed.

In the framework in which we have calculated the $\Delta\phi$ \dis\ 
of the associated particles, it is implied that the \pt \dis\ of the 
pions will have an inverse slope $T=0.33$ GeV for $1.5<p_T<3$ GeV/c 
and that the $p/\pi$ ratio will increase as $0.3\,p_T$ in the same 
region for the same reasons as those found to explain similar 
phenomena related to non-strange triggers \cite{hy,ch}.

While we have focused our attention in this paper on the 
$\Omega$-triggered events, it is natural to predict that exactly the 
same features will be observed for the particles associated with the 
production of $\phi$ under the same conditions. Both $\phi$ and 
$\Omega$ are produced from the same enhanced thermal source, so their 
associated particles should both be formed from the ridges of similar 
characteristics.

\section*{Acknowledgment}
  We are grateful to Jana Bielcikova for her valuable help and 
together with Betty Abelev we thank them for their cooperative 
communication. This work was supported, in part,  by the U.\ S.\ 
Department of Energy under Grant No. DE-FG02-92ER40972.

\newpage
\begin{figure}[htbp]
\centering
\includegraphics[width=6in]{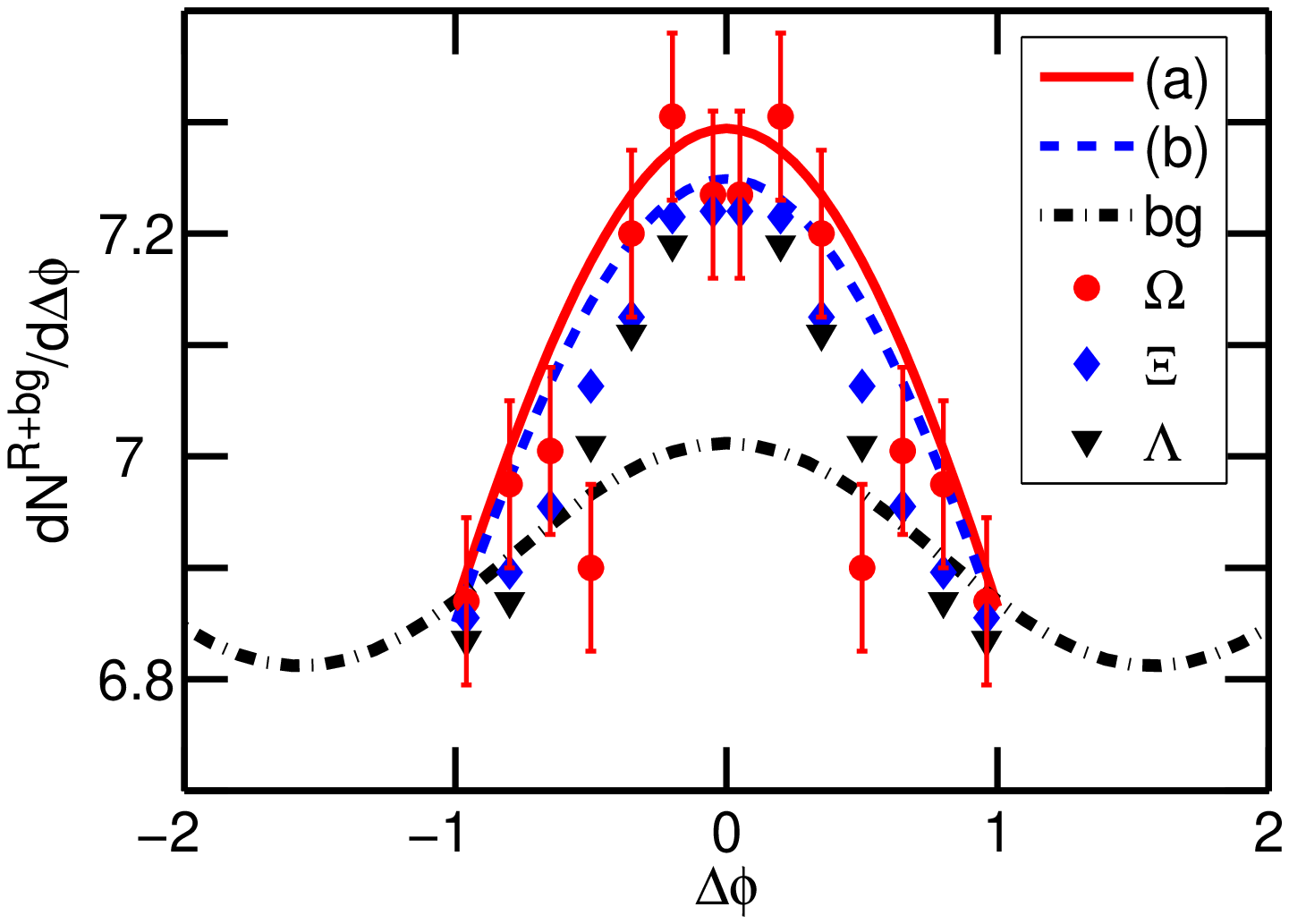}
\caption{
(Color online) Calculated associated particle distributions in events triggered by $\Omega$ for (a) solid (red) line, $H_0=0.795$, and (b) dashed (blue) line, $H_0=0.790$. The data points are from \cite{jbs} for three hyperon triggers. The dashed-dotted line is the background.
}
\end{figure}

\newpage
\begin{figure}[htbp]
\centering
\includegraphics[width=5in]{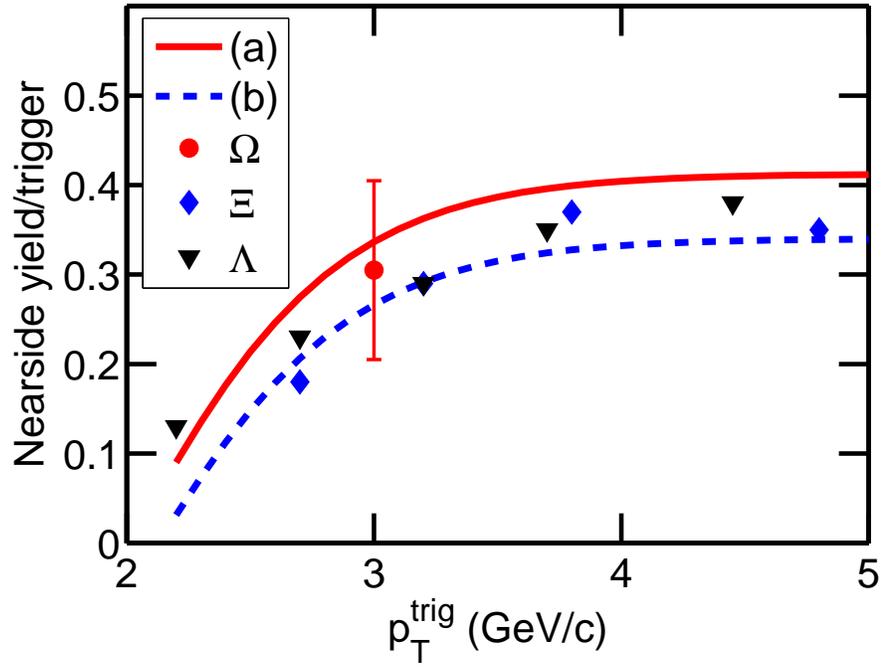}
\caption{(Color online) The yield of the associated particles on the 
near side for $\Omega$ trigger as a function of the trigger momentum. 
The solid, dashed lines and data points are as in Fig.\ 1.}
\end{figure}

\end{document}